\documentclass[english]{article}
\usepackage[latin9]{inputenc}
\usepackage{mathrsfs}
\usepackage{amsmath}
\usepackage{amssymb}
\usepackage{amsthm}
\usepackage{graphicx}
\usepackage{babel}
\usepackage{authblk}
\usepackage{hyperref}
\usepackage{url}
\usepackage{xcolor}

\title{\LARGE Parallel Magnetic Resonance Imaging \\
as Approximation in a Reproducing Kernel Hilbert Space}

\DeclareMathOperator*{\argmin}{argmin}

\author{Vivek Athalye}
\author{Michael Lustig}
\author{Martin Uecker}

\date{August 12, 2014}

\affil{Department of Electrical Engineering and Computer Sciences, University of California, Berkeley, CA 94720}
\affil{Correspondence to: \protect\url{uecker@eecs.berkeley.edu}}

\begin{document}
\maketitle

\global\long\def\xb{\boldsymbol{x}}
\global\long\def\yb{\boldsymbol{y}}
\global\long\def\rb{\boldsymbol{r}}
\global\long\def\fb{\boldsymbol{f}}
\global\long\def\Kb{\boldsymbol{K}}
\global\long\def\Rb{\boldsymbol{K}} 
\global\long\def\gb{\boldsymbol{g}}
\global\long\def\ub{\boldsymbol{u}}
\global\long\def\wb{\boldsymbol{w}}
\def\enc{\epsilon}
\def\hfb{\boldsymbol{\hat f}}

\section*{Abstract}

In Magnetic Resonance Imaging (MRI) data samples are collected
in the spatial frequency domain (k-space), typically
by time-consuming line-by-line scanning on a Cartesian grid. 
Scans can be accelerated by simultaneous acquisition of data using 
multiple receivers (parallel imaging), and by using
more efficient non-Cartesian sampling schemes.
To understand and design k-space sampling patterns, a theoretical 
framework is needed to analyze how well arbitrary sampling patterns
reconstruct unsampled k-space using receive coil information.
As shown here, reconstruction from samples at arbitrary locations
can be understood as approximation of
vector-valued functions from the acquired samples and formulated
using a Reproducing Kernel Hilbert Space (RKHS) with a matrix-valued
kernel defined by the spatial sensitivities of the receive coils.
This establishes a formal connection between approximation theory and
parallel imaging. Theoretical tools from approximation theory
can then be used to understand reconstruction in k-space and to
extend the analysis of the effects of samples selection beyond the
traditional image-domain g-factor noise analysis 
to both noise amplification and
approximation errors in k-space. 
This is demonstrated with numerical examples.

\smallskip
\noindent \textbf{Keywords:} Reproducing Kernel Hilbert Space, Magnetic Resonance Imaging, Image Reconstruction, Approximation, Inverse Problems, Non-Cartesian Sampling

\section{Introduction}

Magnetic Resonance Imaging (MRI) is a non-invasive tomographic imaging
technique with many applications in medicine and biomedical research.
Because it is based on serial scanning of the spatial frequency domain
(known as k-space) by switching of magnetic field gradients, it is
rather time consuming and susceptible to motion artifacts.
Parallel MRI
uses multiple receivers simultaneously
to accelerate the measurement process. Different receive
coils exhibit different spatial sensitivity profiles. 
Because receive coils' different spatial sensitivity profiles provide additional information
about the spatial origin of the signal, the signal can be restored from data which has been
sampled below the Nyquist limit~\cite{SMASH,SENSE,GRAPPA}.

Most naturally, image reconstruction can be formulated as an inverse
problem as in SENSE, where the image is estimated from the acquired data 
by solving a linear signal model~\cite{Ra93,SENSE,CGSENSE,Kannengiesser2000}.
The conditioning of this system depends critically on the number and 
positions of the acquired samples.
Although using local approximations in k-space, the
earlier SMASH method is based on the same fundamental principles as
SENSE.  A classical review of parallel imaging methods and a discussion of 
the relationship between SENSE and SMASH can be found 
in \cite{Sodickson2001}.
The reconstruction from non-Cartesian (scattered) samples 
can also be formulated as an inverse problem and can be solved
exactly in a continuous Hilbert space formulation~\cite{Bertero1985,Walle2000} 
or more commonly using discretization 
and efficient gridding techniques~\cite{SULLIVAN,JACKSON,NUFFT,BEATTY}.
Non-Cartesian sampling can be combined with SENSE~\cite{CGSENSE} and
other advanced reconstruction algorithms
(see \cite{Uecker12,Seiberlich2011,TRACER,GRASP} for recent examples).

Because the reconstruction problem is well-posed
for sufficiently dense and regular sampling, a different two-step
reconstruction strategy is
applied in certain k-space methods.
Using the acquired samples,
a vector-valued function is approximated on a Nyquist-sampled grid
in k-space, which is then transformed to the image domain 
for all coils and only then
combined into a final image. This strategy is used in
coil-by-coil SMASH~\cite{CBCSMASH}, GRAPPA,
and similar methods, {which have first been} formulated for sampling 
on a grid and later extended to non-Cartesian sampling in various 
ways~\cite{PARS,Heidemann2006,Samsonov2006,Beatty2007,Huang2007,SeiberlichGROG,SeiberlichGROG2,Codella,Seiberlich2011}.

Previously, we have shown how both types of reconstruction
constrain the data to a subspace spanned by the spatial sensitivity
profiles of the receive coils~\cite{Uecker13}.
In this work, we extend this idea by showing that parallel imaging 
from arbitrary
- Cartesian as well as non-Cartesian - samples in k-space can be expressed formally as
the approximation problem in a Reproducing Kernel Hilbert Space (RKHS)~\cite{TORKHS,SCDATA}.
A RKHS is a Hilbert space of functions where the point-evaluation
functionals are continuous, {\it i.e.} they are compatible with the norm of the
Hilbert space. 
This is a natural and intuitive property that means functions
close in norm difference are also close at each point and provides 
the additional structure necessary to describe sampling
in a Hilbert space setting. A RKHS is uniquely characterized by its reproducing kernel.
In parallel MRI, the reproducing kernel is determined by the coil sensitivities,
which can be derived directly from the basic signal equation.
While some related ideas can be found in the literature, {\it i.e.}
GRAPPA has been related to the geostatistical framework of Kriging~\cite{HeberleinSeattle06} 
and the ``kernel trick'' known from support vector machines has been used to develop
a non-linear variant of GRAPPA~\cite{NLGRAPPA}, a full mathematical description 
has so far not been available. This gap is closed in the present work by formulating
parallel imaging in the framework of approximation theory. It does
not only provide an optimal interpolation formula as a (theoretical) basis
for image reconstruction in parallel MRI, but also enables a
much deeper understanding of the reconstruction problem itself. In particular,
the {\em power function}~\cite{SchabackPower}
and Frobenius norm maps, 
that naturally come out of the RKHS formulation,
give local bounds of the approximation error 
and about noise amplification
in multi-coil k-space or - with a small extension -
directly for the Fourier transform of the image. 
Both functions depend on the sample
points but not on the 
data and can be used to study the effect of sample selection on the 
reconstruction error. This is demonstrated with numerical examples.

\section{Theory}

\subsection{Overview}

\begin{figure}
\begin{center}
\includegraphics[width=\columnwidth]{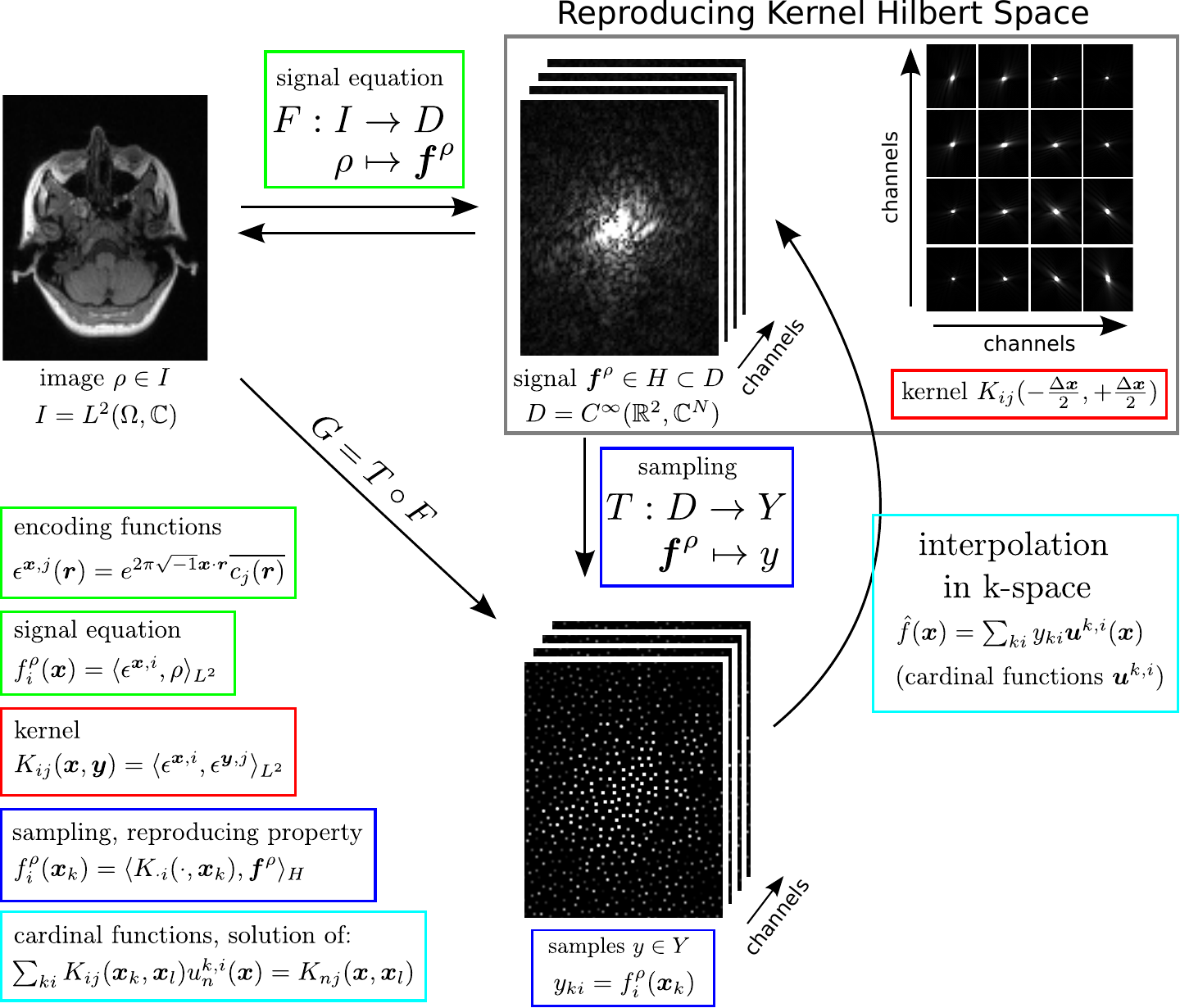}
\end{center}
\caption{Image reconstruction for parallel MRI as approximation in a reproducing kernel Hilbert space.} \label{fig:OVERVIEW}
\end{figure}

An overview of the theory developed in the following
is shown in Figure~\ref{fig:OVERVIEW}. Please refer to 
Appendix~\ref{app:NOTATION} for some comments about the notation
and to Table~\ref{tab:NOTATION} for a list of important symbols.

We consider parallel imaging as an inverse problem with 
a linear forward model $F: I \rightarrow D$,
which maps from a Hilbert space of images~$I$ to a data space~$D$.
The range of $F$ is the space of ideal signals~$H \subset D$.
From the data space a set $y \in Y$ of samples are acquired, which 
is described by a sampling operator $T$. Then, the general setting
is the following:
\begin{align*}
	\begin{array}{cccc}
		I & \xrightarrow{F} & H & \subset D \\
		 & \searrow_G & \downarrow_{T} & \\
		 & & Y &
	\end{array}
\end{align*}
Here, $G = T \circ F$. 
A general formulation of
linear inverse problems with discrete data
can be found in~\cite{Bertero1985}.
The inverse problem can be 
solved by computing a regularized least-squares solution by minimizing a functional
\begin{align} \label{eq:MINI}
	\rho_{\alpha} = G^{\dagger}_{\alpha} y = \argmin_{\hat \rho} \|G \hat \rho - y\|_2^2 + \alpha R(\hat \rho)
\end{align}
with a suitable regularization
term $R$. 
In the limit $\alpha \rightarrow 0$, this yields a minimum-norm least-squares
solution (MNLS).
In general, the mapping $F$ is injective and has a stable 
inverse defined on its range~$H$. Alternatively to solving the inverse problem 
directly, one can approximate a function $\hfb \in H$ from the data~$y \in Y$
and obtain a solution by computing $\hat \rho = F^{-1} \hat \fb$.

\begin{table}
\begin{center}
\begin{tabular}{l|l}
	$\xb, \yb$ & k-space positions \\
	$S$ & finite set of sample locations \\
	$\xb_k, \yb_l$ & indexed sample locations in $S$ \\
	$i,j,n$ & indices used for vector components (channels) \\
	$H$	& Hilbert space of vector-valued k-space functions \\
	$\Rb^{\xb,i}(\yb)$ & representer of evaluation of channel $i$ at $\xb$ \\
	$K_{ij}(\xb, \yb)$ & matrix-valued kernel \\
	$M_{ki,lj}$ & kernel matrix \\
	$\ub^{k,i}(\xb)$ & cardinal functions \\
	$\langle \cdot, \cdot \rangle_H$ & inner product in $H$ \\
	$P_n(\xb)$ & Power function for channel $n$ \\
	$\Omega$ & the field of view (FOV) \\
	$L^2(\Omega, \mathbb{C})$ & square-integrable functions on $\Omega$ \\
	$\rb$ & image-domain position \\
	$c_j(\rb)$ & coil sensitivity map for channel $j$ \\
	$\enc^{\xb,i}(\rb)$ & encoding functions \\
	$\rho(\rb)$ & image \\
	$\overline{(\cdot)}$ & complex conjugate
\end{tabular}
\end{center}
\caption{Important symbols.} \label{tab:NOTATION}
\end{table}

\subsection{Parallel Magnetic Resonance Imaging}

We begin with the standard setup of the parallel imaging problem.
For concreteness, we consider two-dimensional imaging.
Let the magnetization image $\rho:\mathbb{R}^{2}\rightarrow\mathbb{C}$
belong to the space $L^{2}(\Omega, \mathbb{C})$ of square-integrable functions
with compact support~$\Omega$ on a subset of the plane called the field of view
(FOV).
The forward operator $F$ maps magnetization images to smooth signals in k-space:
\begin{align}
	F: L^2(\Omega, \mathbb{C}) & \rightarrow C^{\infty}(\mathbb{R}^2, \mathbb{C}^N) \\
		\rho & \mapsto \fb^{\rho} = F \rho
\end{align}
Each vector component, which is the signal of one of $N$ receive coils, is
given by the signal equation:
\begin{equation}\label{eq:SIGNAL}
	f_{j}^{\rho}(\boldsymbol{x}) = \int_{\Omega}d\boldsymbol{r}\, \rho(\boldsymbol{r}) {c_{j}(\boldsymbol{r}) e^{-2\pi\sqrt{-1}\xb \cdot \boldsymbol{r}}}\quad\quad\quad1\le j\le N
\end{equation}
In words, the $j^\mathrm{th}$ component of the vector-valued function $\fb^{\rho}$
is the Fourier transform of coil $j$'s image.
The k-space signals are smooth because they are Fourier transforms of compactly supported functions.
The coil sensitivities $c_{j}$ are generally smooth, complex-valued
functions in image space describing the spatial sensitivity
profiles of each receiver coil. 
In areas where all coil sensitivities vanish
simultaneously, no information about the image can be recovered.
Without loss of generality, we will simply assume that the definition 
of $\Omega$ excludes such areas.
Using the inner product definition
\begin{align}
	\langle \rho, \sigma \rangle_{L^2} = 
	\int_{\Omega}d\boldsymbol{r}\, \overline{\rho(\boldsymbol{r})} \sigma(\boldsymbol{r}) 
\end{align}
and the {\em encoding functions} $\enc^{\xb,j}(\rb) = e^{2\pi\sqrt{-1}\xb \cdot \rb} \overline{c_{j}(\rb)}$~\cite{SENSE},
we can write $f_j^{\rho}(\xb) = \langle \enc^{\xb, j}, \rho \rangle_{L^2}$.
During the measurement process samples at a finite number of locations
$\boldsymbol{x}_k \in S \subset \mathbb{R}^2$ are collected.
Samples can be assumed to be corrupted by i.i.d. complex Gaussian 
white noise. Although in practice receive channels might have different 
noise levels and correlations, this can be removed by a 
prewhitening step and a change-of-variable transformation
of the coil sensitivities~\cite{CGSENSE}.

\subsection{Reproducing Kernel Hilbert Space}

The vector-valued functions considered in parallel imaging have the
particular structure specified in Equation~\ref{eq:SIGNAL}. We now
encapsulate this structure within a reproducing kernel Hilbert space $H$
with a matrix-valued kernel~\cite{Narcowich,Fuselier}.

Let $X$ be a set of points, and $H$ a Hilbert space of vector-valued
functions on $X$. $H$ is an RKHS if the point-evaluation functionals
$L^{\xb}:H\rightarrow \mathbb{C}^{N}$, $\boldsymbol{f}\mapsto\boldsymbol{f}(\boldsymbol{x})$
are continuous for all $\boldsymbol{x} \in X$.
From the Riesz representation theorem, it then follows
that there are unique functions $\Rb^{\boldsymbol{x},i}\in H$
for each $\xb\in X$ and each vector component $1\le i\le N$ such
that $f_{i}(\xb)=\left\langle \Rb^{\xb,i},\fb\right\rangle_H$.
As before, we define the inner product $\langle \cdot, \cdot \rangle_H$ in $H$
to be conjugate linear in the first argument.
The functions $\Rb^{\xb,i}$ are called representers of evaluation.
They span $H$, but do not generally form a basis.
If a series of functions $\fb^n$ converges to $\fb^{\star}$ 
in the Hilbert space norm, then for any $\boldsymbol{x} \in X$
and vector component $1\le i\le N$, we have 
\begin{align}
	|f_i^n(\xb) - f^{\star}_{i}(\xb)| & = |\left\langle \Rb^{\xb,i},\fb^n - \fb^{\star} \right\rangle_H| \nonumber\\
		& \leq \| \Rb^{\xb,i} \|_H \| \fb^n - \fb^{\star} \|_H~.
\end{align}
This means that convergence in norm implies point-wise convergence
and that local bounds can be obtained using information about the representers.

The structure of the space can then be described by a positive-definite
matrix-valued kernel $K: X\times X\rightarrow\mathbb{C}^{N\times N}$
such that an element of the kernel
$K_{ij}(\xb,\yb) = \left\langle \Rb^{\xb,i},\Rb^{\yb,j}\right\rangle_{H}$.
A RKHS is uniquely characterized by its positive-definite kernel
and to every positive-definite kernel there is a unique RKHS.
From the definition of the representers of evaluation it follows 
that $K_{ij}(\xb,\yb) = K_{i}^{\yb,j}(\xb)$,
the $i^\mathrm{th}$ component of the vector-valued function $\Rb^{\yb,j}$
that evaluates the $j^\mathrm{th}$ component of a function at $\yb$.
Thus, the {\em reproducing property} holds:
\begin{align} \label{eq:REPRO}
f_{j}(\yb) & = \left\langle {K}_{\cdot j}(\cdot,\yb), \fb\right\rangle_H \qquad \forall \yb \in X
\end{align}

Applying this framework to parallel MRI, the space $H$ is the range of $F$, {\it i.e.} it consists of ideal signals 
$\fb$ on $\mathbb{R}^2$
given by Equation~\ref{eq:SIGNAL} for all possible images $\rho\in L^{2}(\Omega, \mathbb{C})$.
We can assume that at least one of the coil sensitivities $c_{j}$ is non-zero at
each point $\rb\in \Omega$. 
Then $F$ can be inverted by applying an inverse
Fourier transform (practical computation can be done on a Nyquist-sampled
grid) and dividing by a non-zero $c_{i}$. This enables the following definition
of an inner product in $H$:
\begin{equation}
	\left\langle \fb,\gb\right\rangle_H := \langle F^{-1}\fb, F^{-1}\gb\rangle_{L^2}
\end{equation}
With this inner product, we formulate our main result: \\

\noindent
$\textit{Theorem}$: 
The space $H$ of ideal multi-channel signals in MRI is a RKHS with kernel:
\begin{align}\label{eq:KERNEL}
	K_{ij}(\xb,\yb) & = \langle \enc^{\xb, i}, \enc^{\yb, j} \rangle_{L^2} \nonumber\\
	 	        & = \int_{\Omega}d\rb \, e^{-2\pi\sqrt{-1}(\xb-\yb) \cdot \rb}c_{i}(\rb)\overline{c_{j}(\rb)}
\end{align}
\begin{proof}
We must show two properties: For each $\yb\in \mathbb{R}^2$
and $1\le j\le N$ the $\Rb^{\yb,j}$ must lie in $H$,
and $\left\langle \Rb^{\yb,j},\fb\right\rangle_H =f_{j}(\yb)$.
We observe that the kernel can be obtained by 
applying $F$ to the encoding functions $\enc^{\yb,j}$:
\begin{align}
 \left( F \enc^{\yb,j} \right)_i(\xb) 
	= \langle \enc^{\xb, i}, \enc^{\yb, j} \rangle_{L^2} 
	= K_{ij}(\xb, \yb) 
 	= K^{\yb,j}_i(\xb) 
\end{align}
Thus, $\Rb^{\yb,j}$ is in the range of $F$ and 
therefore in $H$. The second property follows directly from the definition
of the inner product. For every $\fb^{\rho} = F\rho$, it follows:
\begin{align}
	\left\langle \Rb^{\yb,j}, \fb^{\rho}\right\rangle_H 
	= \left\langle F \enc^{\yb,j}, F \rho\right\rangle_H 
	= \left\langle \enc^{\yb,j}, \rho\right\rangle_{L^2} 
	= f_{j}^{\rho}(\yb)
\end{align}
\qedhere
\end{proof}

In words, $K_{ij}(\xb,\yb)$ captures the similarity between encoding functions
$ \enc^{\xb, i}$ and $\enc^{\yb, j}$. Note that $K_{ij}(\xb,\yb) = K_{ij}(\xb + \Delta,\yb + \Delta)$,
and so the kernel is shift invariant.
It should be noted that there is some freedom in the choice 
of the inner product. Using a different inner product will
lead to a different kernel. The inner product used here
corresponds to the inner product of the Hilbert space of 
images. As shown later, the final reconstruction will then be 
optimal with respect to this norm.

Having characterized the multi-channel k-space of ideal signals
as an RKHS with the shift-invariant kernel given in 
Equation~\ref{eq:KERNEL}, we can
now proceed to describe sampling and reconstruction in this
framework.

\subsection{Sampling and Reconstruction}

Sampling and reconstruction from arbitrary samples can be described in 
the framework of approximation theory (see \cite{SCDATA} as 
general reference). Because it is usually
formulated for the scalar case only, we summarize the main results using 
our notation.

Samples are collected at a finite number of locations
$\xb_k \in S \subset \mathbb{R}^2$. Ideal samples of $\fb \in H$ are then given by 
the inner product evaluations $\fb_{i}(\xb_k)=\left\langle \Rb^{\xb_k,i}, \fb\right\rangle_H$ 
for all $i \in 1,\cdots,N$ and $\xb_k \in S$. Assuming
no measurement error,
a solution to the reconstruction problem is usually defined as the
function $\hfb \in H$ of smallest norm which interpolates $\fb$ at the
sample locations, {\it i.e.} $\hfb(\xb_l) = \fb(\xb_l)$.
We first define a measurement subspace $H_{S}\subset H$ that is spanned 
by $\Rb^{\xb_{k},i}$ for all $\xb_{k}\in S,\:1\le i\le N$. Thus,
any function $\fb \in H_S$
can be represented as
\begin{align}
	\fb = \sum_{k = 1}^{|S|}\sum_{i=1}^{N}a_{k,i}\Rb^{\xb_{k},i}~.
\end{align}
This subspace turns out to be the right space for interpolation: In the
absence of errors, all
functions in $H_{S}$ can be recovered exactly (see below), while the functions
 $\fb^{\perp} \in H_{S}^{\perp}$ are those for which the samples provide no information, i.e.
$\fb^{\perp}(S) = \{ 0 \}$ (by Eq.~\ref{eq:REPRO}).
The minimum-norm interpolant $\hfb$ for $\fb \in H$ for the
samples $S$ is the projection~$\fb^{\parallel}$ of $\fb$ onto $H_{S}$. 
To compute this projection, we could directly solve for coefficients $a_{k,i}$ such
that ${\hfb}(\xb_k) = \fb(\xb_k)$ for $\xb_{k} \in S$.

Instead, we formulate
a generic solution which only depends on the sample locations and not on the data.
The {\em kernel matrix} $M \in \mathbb{C}^{|S|N\times |S|N}$ is constructed
by evaluating the kernel at all sample positions $\xb_k \in S$, {\it i.e.}
$M_{ki,lj} = K_{ij}(\xb_k, \xb_l)$. Here, the pairs of indices $k$, $i$ 
and $l$, $j$ have each been combined to obtain the two indices of the matrix.
For each point $\xb$, interpolation weights called {\em cardinal functions} $\ub^{k,i}(\xb)$ 
can then be computed by solving a linear system of equations:
\begin{align}\label{eq:CARDINAL}
	\sum_{k = 1}^{|S|} \sum_{i=1}^N M_{ki,lj} \, \ub^{k,i}(\xb) = \Rb^{\xb_l, j}(\xb)
\end{align}
The cardinal functions are independent from $\fb$ and interpolate
functions from $H$ according to (see Appendix~\ref{app:EXACT})
\begin{align}\label{eq:INTERPOL}
	\boldsymbol{\hat f}(\xb) & = \sum_{k = 1}^{|S|} \sum_{i=1}^N f_i(\xb_k) \ub^{k,i}(\xb)~.
\end{align}

In principle, parallel MRI reconstruction can 
be performed using this formula: The 
interpolation formula can be used to compute samples on a Nyquist-sampled
grid followed by an FFT algorithm to compute an image for each coil.
Evaluation of Equation~\ref{eq:CARDINAL} requires the solution of
the linear system of equations of size $|S|N \times |S|N$ for each point
of the Nyquist-sampled grid. 
Although all solutions can be obtained efficiently after computation
of the pseudo-inverse (or Cholesky decomposition) of the kernel matrix,
this is still too expensive for image 
reconstruction in clinical applications due to the size of this matrix. 
Nevertheless, this formula is closely related to more efficient practical
algorithms such as GRAPPA, which are based on an approximation.

\subsection{Approximation Error and Noise}

A useful tool to estimate approximation errors is the 
{\em power function}~$\boldsymbol{P}$~\cite{SchabackPower}.
It yields point-wise bounds of the approximation error:
\begin{align}
	 | \fb_n(\xb) - \hfb_n(\xb) |^2 \leq \| \fb \|_H^2 \cdot P^2_n(\xb)
\end{align}
The power function depends only on the sample locations and is
independent from the data values. It is given in terms of the
kernel and the cardinal functions (see Appendix~\ref{app:BOUND}).
This concept enables us to understand how good our sample set $S$ is
for approximating $\fb$ at unacquired points. A single
combined power function can be obtained as the root of sum of squares
of the power functions for all coils.
A large power function indicates that we should expect a large approximation error
at this point. A small value of the
power function means that the function is approximated well
from the available samples and that a new sample at $\xb$
would not provide much information.

Another quantity of interest is the
Frobenius norm of the local reconstruction operator:
\begin{align}
	n_n(\xb) = \sqrt{ \sum_{k=1}^{|S|} \sum_{i=1}^N \left|\ub^{k,i}_n(\xb)\right|^2 }
\end{align}
It relates to the stability of the interpolation,
describing the local noise amplification in k-space
for each individual coil (or for all coils using a combined map).
This yields different and complementary information to the g-factor maps
in the image domain~\cite{SENSE,BREUERGF}. While the g-factor map yields
information about how noise affects the final reconstructed image,
these new noise amplification maps yield useful information
about the source of the noise in k-space and can guide the design
of optimal sampling patterns.
A related  quantity is the classical {\em Lebesgue function},
which is defined as
\begin{align}
	l_n(\xb) = \sum_{k = 1}^{|S|} \sum_{i=1}^N \left| \ub^{k,i}_n(\xb) \right| ~.
\end{align}
Its maximum is the {\em Lebesgue constant}, which
can be used to bound the interpolation error in the maximum norm when
the error of the data is bounded (for example, see \cite{Trefethen2013}).

\subsection{Minimum-Norm Reconstruction}\label{app:MNLS}

Eq.~\ref{eq:MINI} defines the MNLS solution of SENSE in the 
continuous (not discretized) space of image.
It is more precise than a conventional SENSE reconstruction
using dirac distributions as voxel functions, which is
affected by truncation artifacts~\cite{Yuan2006}.
A MNLS reconstruction in the image domain can be computed 
directly~\cite{Sodickson2001} or approximated efficiently by computation 
on an image-domain grid with higher resolution~\cite{Tsao2003,Gonzales2006,Uecker09}.

With our choice of the inner product, the
k-space recovered with Eq.~\ref{eq:INTERPOL} corresponds 
to this MNLS reconstruction of the image.
Solving for the cardinal functions (Eq.~\ref{eq:CARDINAL}) 
using the pseudo-inverse $M^{\dagger}$
of the kernel matrix and
inserting into the interpolation formula (Eq.~\ref{eq:INTERPOL}) yields:
\begin{align}
	\hat \fb = \sum_{l = 1}^{|S|} \sum_{j=1}^N \Rb^{\xb_l, j} \sum_{k = 1}^{|S|} \sum_{i=1}^N  M^{\dagger}_{lj,ki} f_i(\xb_k) 
\end{align}
Expanding the relations
\begin{align}
	M_{ki,lj} & = K_{ij}(\xb_k, \xb_l) = \langle \enc^{\xb_k, i}, \enc^{\xb_l, j} \rangle_{L^2} \\
	\Rb^{\xb_l, j} & = F \enc^{\xb_l, j} 
\end{align}
and noting that the result can be re-written using the 
forward operator 
\begin{align}
	G: L^2(\Omega, \mathbb{C}) & \rightarrow \mathbb{C}^{|S|N} \\
		\rho & \mapsto y_{ki} = \langle \enc^{\xb_k, i}, \rho \rangle_{L^2} 
\end{align}
and its adjoint $G^H$, one obtains $M = G G^H$ and further
\begin{align}
	  \hat \fb &= F G^H \left( G G^H \right)^{\dagger} y \\
	  &= F G^{\dagger} y~.
\end{align}
From this formulation it is directly evident that the k-space
signal interpolated with Eq.~\ref{eq:INTERPOL} is the same as the
signal predicted with the operator $F$ from the MNLS
solution  $G^{\dagger} y$ of the continuous (not discretized
 in the image domain) 
SENSE problem. No discretization errors arise, because the kernel 
matrix $M$, {\it i.e.} \(G G^H\), is a finite-dimensional matrix 
even for the continuous case. Of course, this is not a unique feature 
of the proposed formulation: The relation $G^H \left( G G^H \right)^{\dagger} = G^{\dagger}$
used in the last equation can be applied in reverse to the SENSE
problem to directly compute the MNLS solution without discretization 
errors~\cite{Sodickson2001}. The same idea has also been proposed 
earlier for the reconstruction of non-Cartesian data from a single 
coil~\cite{Walle2000}.

\subsection{Relationship to GRAPPA}

Equation~\ref{eq:INTERPOL} is similar to the reconstruction
part of the GRAPPA algorithm, with the {\em GRAPPA weights} corresponding
to the {\em cardinal functions}. To make actual computation feasible,
GRAPPA uses only samples in a small patch near a given point.
If $S_{\xb} \subset S$ is the set of samples near $\xb$ which are
used for reconstruction, then the GRAPPA reconstruction~$\hfb_{G}$
is given by:
\begin{align}
	\hfb_{G}(\xb) = \sum_{k|\xb_k \in S_{\xb}} \sum_{i=1}^N f_i(\xb_k) \wb^{k,i}(\xb) 
\end{align}
The weights are determined with a calibration procedure. The calibration
(and reconstruction) in the original GRAPPA method is restricted to samples
on a Cartesian  grid. Using shift invariance, the weights~$\wb^{k,i}(\xb)$ are learned
by a least-squares fit at many different grid positions~$\xb_t \in C$ in a calibration
region where all required samples (on the grid) have been acquired:
\begin{align}
w_j^{k,i}(\xb) = \argmin_{\hat w^{k,i}} \sum_{t=1}^{|C|} \Biggl| \sum_{k|\xb_k \in S_{\xb}}\sum_{i=1}^N  f_i(\xb_t + \underbrace{\xb_k - \xb}_{\Delta \xb_k}) \hat w^{k,i} - f_j(\xb_t) \Biggr|^2 
\end{align}
The distance vectors $\Delta \xb_k$ which appear in the sum for a specific target position
$\xb$ depend on the local sampling pattern. 
The least-squares problem can be solved explicitly using the normal equations.
Because in GRAPPA only neighboring samples are used, it is useful
to define a so-called {\em calibration matrix}, which is computed by sliding 
a small window through the calibration area and taking each patch as a row, {\it i.e.}
$A_{t,ki} = f_i(\xb_t + \Delta \xb_k)$ for all possible distance vectors $\Delta \xb_k$
in a small patch. Assuming a special index $0$ with $\Delta \xb_0 = \bf 0$, 
{\it i.e.} $A_{t,0i} = f_i(\xb_t)$, the normal equations 
are given by (with $l$ such that $\Delta \xb_l + \xb \in S_{\xb}$ and $j = 1 \cdots N$)
\begin{align}
	\sum_{k |\Delta \xb_k + \xb \in S_{\xb}} \sum_{i=1}^N \sum_{t=1}^{|C|} \overline{A_{t,lj}} A_{t,ki} \, w_h^{k,i}(\xb) 
		&= \sum_{t=1}^{|C|} \overline{A_{t,lj}} A_{t,0h}~.
\end{align}
Again, this relates to the present framework as 
this is Equation~\ref{eq:CARDINAL} expressed using relative distances
and with a kernel matrix $M_{ki,lj} = \sum_t A_{t,ki} \overline{A_{t,lj}}$.
This kernel matrix is related to an estimate of 
a truncated {\em covariance function} given by (assuming vanishing mean):
\begin{align}
	Cov_{ij}(\Delta \xb) = \frac{1}{|C|} \sum_{t=1}^{|C|} {f_i(\xb_t)}\overline{f_j(\xb_t + \Delta \xb)}
\end{align}

Note, that due to the kernel's shift invariance,
the GRAPPA weights (or cardinal functions) in a local approximation 
of the interpolation formula (Eq.~\ref{eq:INTERPOL}) are identical  
at positions in k-space which share the same local sampling pattern.
In this way, computation time for (quasi-)periodic sampling can be
reduced considerably.

In summary, the GRAPPA algorithm can be re-formulated and
understood as function approximation in a RKHS. 
Instead of the optimal kernel, which has analytically been derived 
from the coil sensitivities in Equation~\ref{eq:KERNEL}, a different kernel 
(which corresponds to a different RKHS) related to the empirical covariance 
function is used. Reconstruction differs from the optimal formula 
by only using local samples for interpolation.

\section{Numerical Examples}

\begin{figure}
\begin{center}
\includegraphics[width = 1\columnwidth]{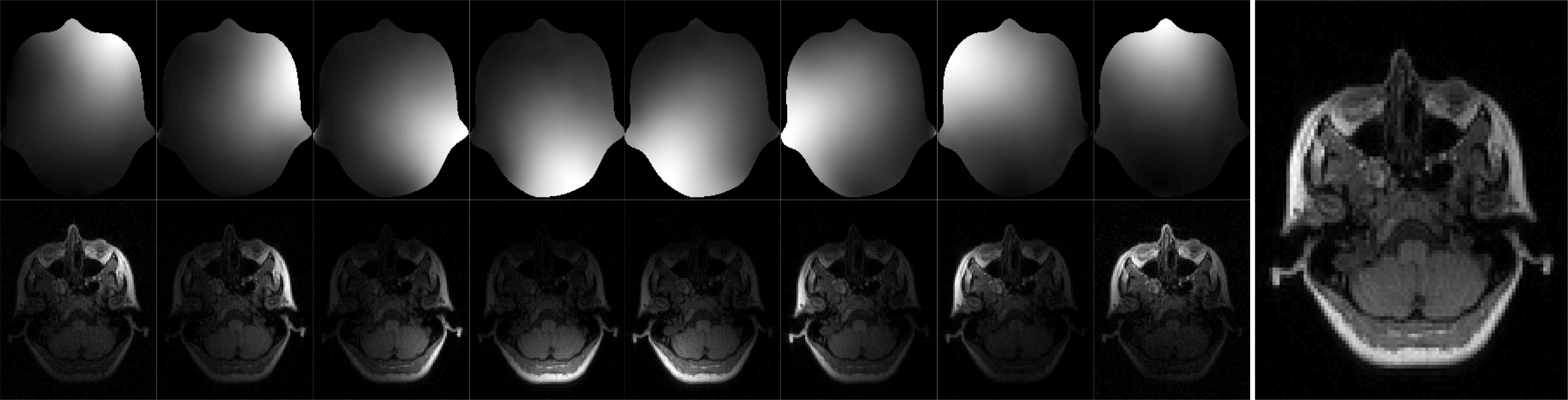}
\caption{Top row: Spatial sensitivity maps for each channel of the receive coil (limited to the support of the object). 
Bottom row: Corresponding coil images computed by Fourier transform of each channel of the fully-sampled data.
Right: Combined image computed as the pixel-wise root of the sum of absolute squares of all coil images.}
\label{fig:SENS}
\end{center}
\end{figure}

\subsection{Methods}

Numerical experiments have been performed using a 2D slice
of size $115 \times 90$ extracted from a larger fully-sampled
3D data set of a human head acquired at 1.5~T using 
an eight-channel head coil (inversion-recovery prepared RF-spoiled
3D-FLASH, TR/TE = 12.2/5.2~ms, TI = 450~ms, FA =~20$^\circ$, BW = 15~KHz).
In order to demonstrate k-space interpolation 
from 
Cartesian and
arbitrary, non-Cartesian samples, sampling patterns were generated 
on a k-space grid oversampled by $3\times$ in each dimension by zero-padding
in the image domain. From this data set samples have been obtained
using Cartesian, Poisson-disc, and uniform random sampling.
Cartesian sampling used an undersampling of
four in one phase-encoding direction ($4 \times 1$) and
of two in both phase-encoding directions ($2 \times 2$).
In addition, different CAIPIRINHA~\cite{BreuerCAIPI} patterns have been
studied. 
The Poisson-disc radius used yielded 2494 samples corresponding 
to an acceleration factor about four relative to the original number
of samples, and the uniform random sampling pattern was generated 
to match this number of samples.

Coil sensitivities have been estimated using ESPIRiT~\cite{Uecker13},
which yields very accurate sensitivities up to point-wise normalization
{and a mask which defines the area with signal.}
From these coil sensitivities the kernel has been computed by evaluating
Equation~\ref{eq:KERNEL} using a zero-padded Fourier transform.
To demonstrate reconstruction errors from interpolation errors only,
{\it i.e.} excluding errors from noise, a synthetic data set was created:
The fully-sampled data was combined into a single image and then
data was {simulated} using the coil sensitivities.

For all sampling patterns, a kernel matrix has been constructed
by evaluating the kernel at the sample positions. 
Especially when some samples are close together as in the
random sampling pattern, corresponding rows and columns in the
kernel matrix are very similar, and the condition number is large.
For this reason, the inversion of the kernel matrix has to be stabilized
with Tikhonov regularization in the presence of noise and numerical errors.
The maximum eigenvalue as determined by power iteration of the kernel 
matrix was $51.0062$ for Cartesian $4 \times 1$, $31.033$ for 
Cartesian $2 \times 2$, $33.863$ for Poisson-disc, and $116.45$
for random sampling. For the CAIPIRINHA patterns the values 
were between $30.0936$ and $35.3207$.
To avoid a large influence on the solution, 
Tikhonov regularization (ridge regression) was used with a much 
smaller parameter of $0.01$ for all experiments.

For each point on an oversampled and extended Cartesian grid, the
cardinal and power functions have been computed. 
To reduce computation time, Equation~\ref{eq:CARDINAL} is solved by forward and
backward substitution using a Cholesky decomposition of the kernel matrix.
To estimate interpolation error and noise amplification at each
position, a combined power function for all coils is
computed by root of sum of squares and the cardinal functions
are combined in the same way which yields the Frobenius norm
of the local interpolation operator.
Using the cardinal function, a Nyquist-sampled k-space has then
been approximated from the acquired samples for synthetic and noisy 
data and transformed into coil images using an FFT.

The simulations were performed using Matlab (The MathWorks, Inc., Natick, MA)
on a cluster with two quad-core
Intel Xeon E5550 CPUs (2.67 GHz) per node. Computing the kernel
matrix of size $\sim20000^2$ 
and  calculating the Cholesky decomposition took about two CPU hours. 
Solving for the cardinal functions for all interpolation points was broken 
into smaller parallel jobs to reduce runtime and memory use. 
To interpolate from the samples to the full $115 \times 90$ grid 
took $\sim350$ CPU hours and to interpolate over
the $3\times$ oversampled and extended grid of size $405 \times 330$
took about $\sim4500$ CPU hours.
Utilizing about $20$ parallel jobs on the cluster, this
took about $6$ hours for the full grid and about $3$ days for the 
oversampled grid.
All computations used double-precision floating-point arithmetic.

\subsection{Results}

\begin{figure}
\begin{center}
\includegraphics[width=1\columnwidth]{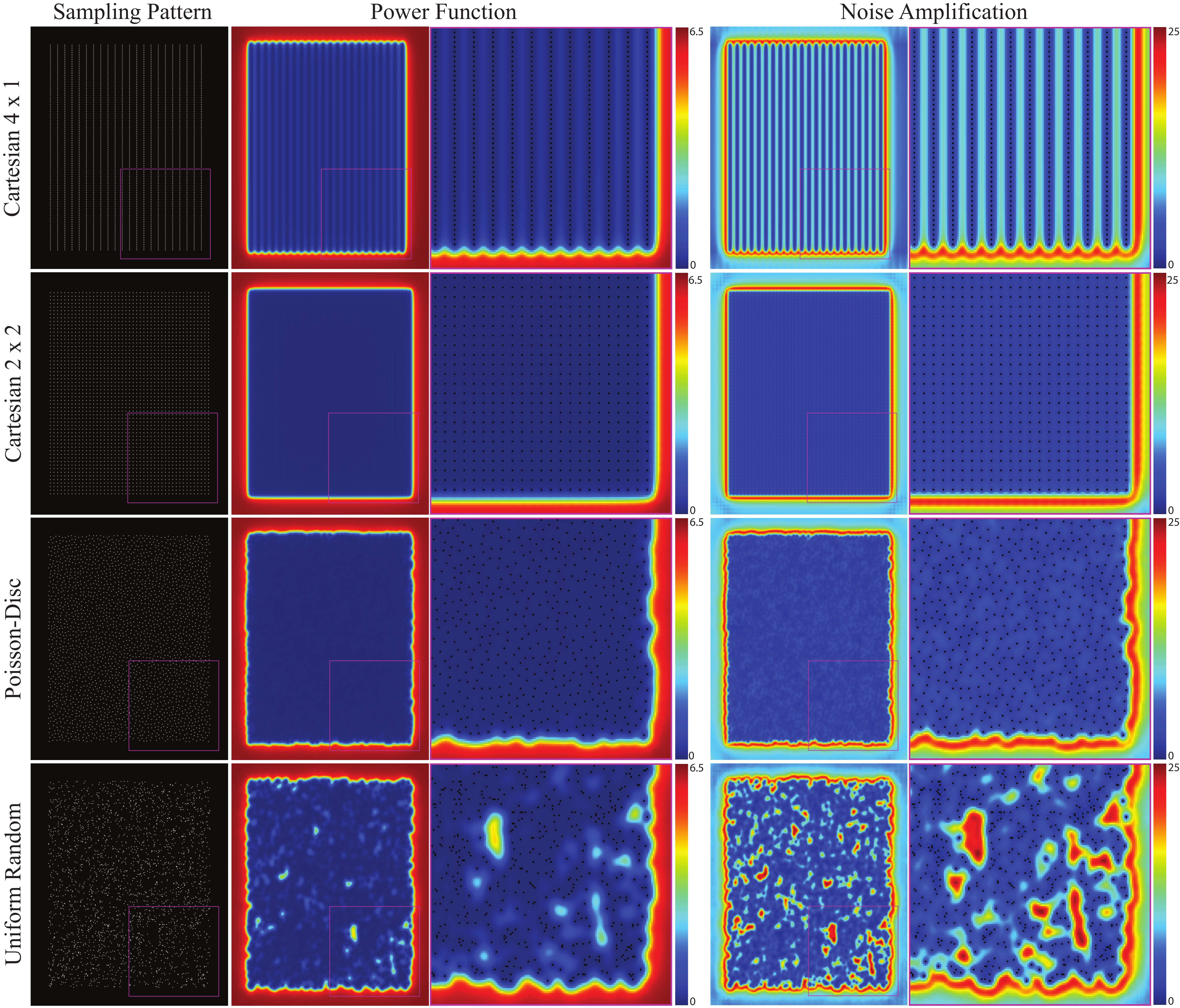}
\caption{Sampling pattern, combined power function, and local noise amplification 
for Cartesian, Poisson-disc, and random
sampling on an oversampled and extended grid. 
The maximum possible value for the power function is $6.5530$
in regions where no information is available. In the blown-up region of the
lower-right corner sample positions are indicated by black dots.}
\label{fig:POWER}
\end{center}
\end{figure}

\begin{figure}
\begin{center}
\includegraphics[width=1\columnwidth]{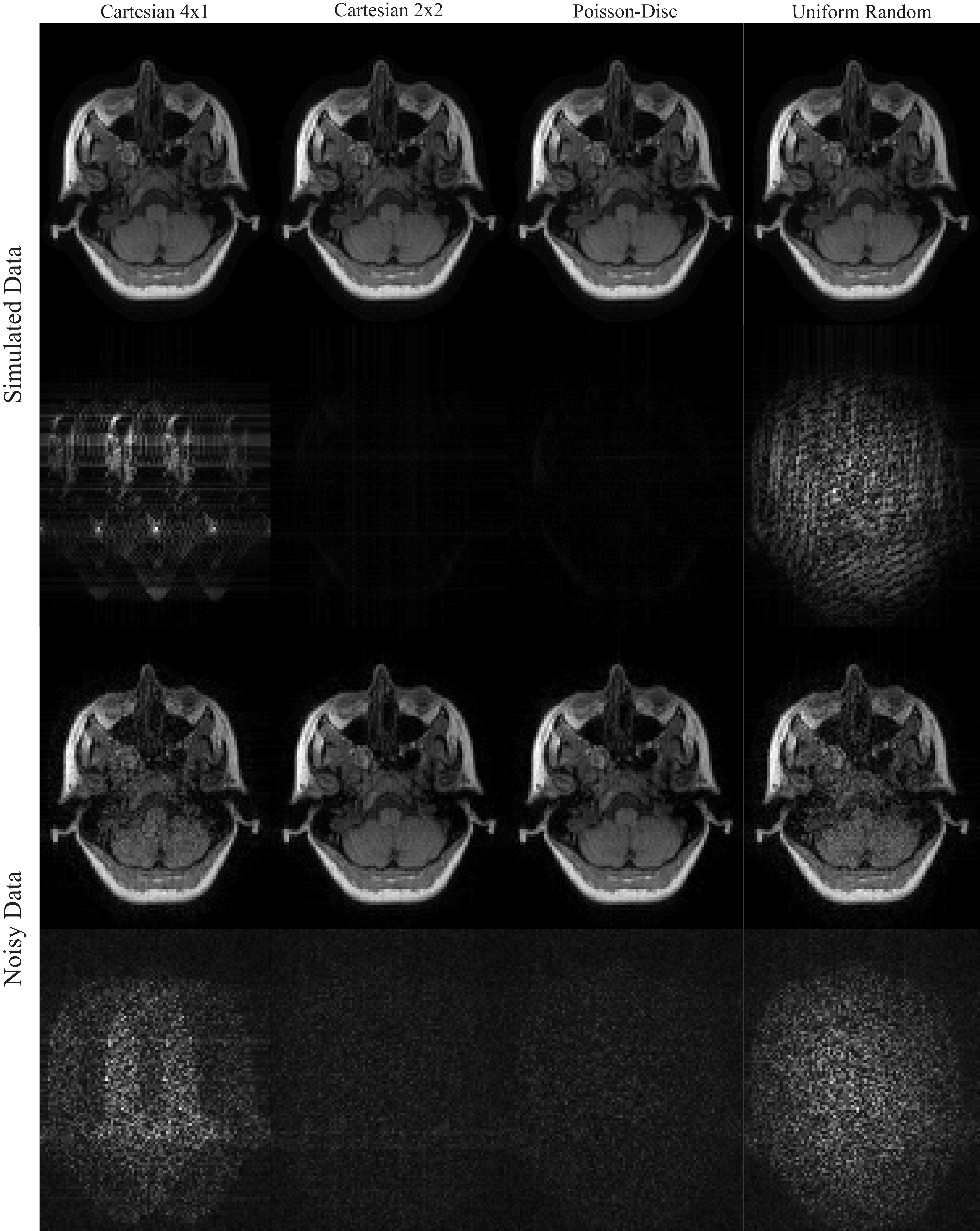}
\caption{Reconstructed images and corresponding error maps for Cartesian, Poisson-disc,
and uniform random sampling for simulated (top) and noisy data (bottom). All sampling schemes used an 
undersampling factor of 4. Error maps have been scaled by a factor of 40 (simulated) and 4 (noisy) to aid visibility.}
\label{fig:RECO}
\end{center}
\end{figure}

\begin{figure}
\begin{center}
\includegraphics[width=1\columnwidth]{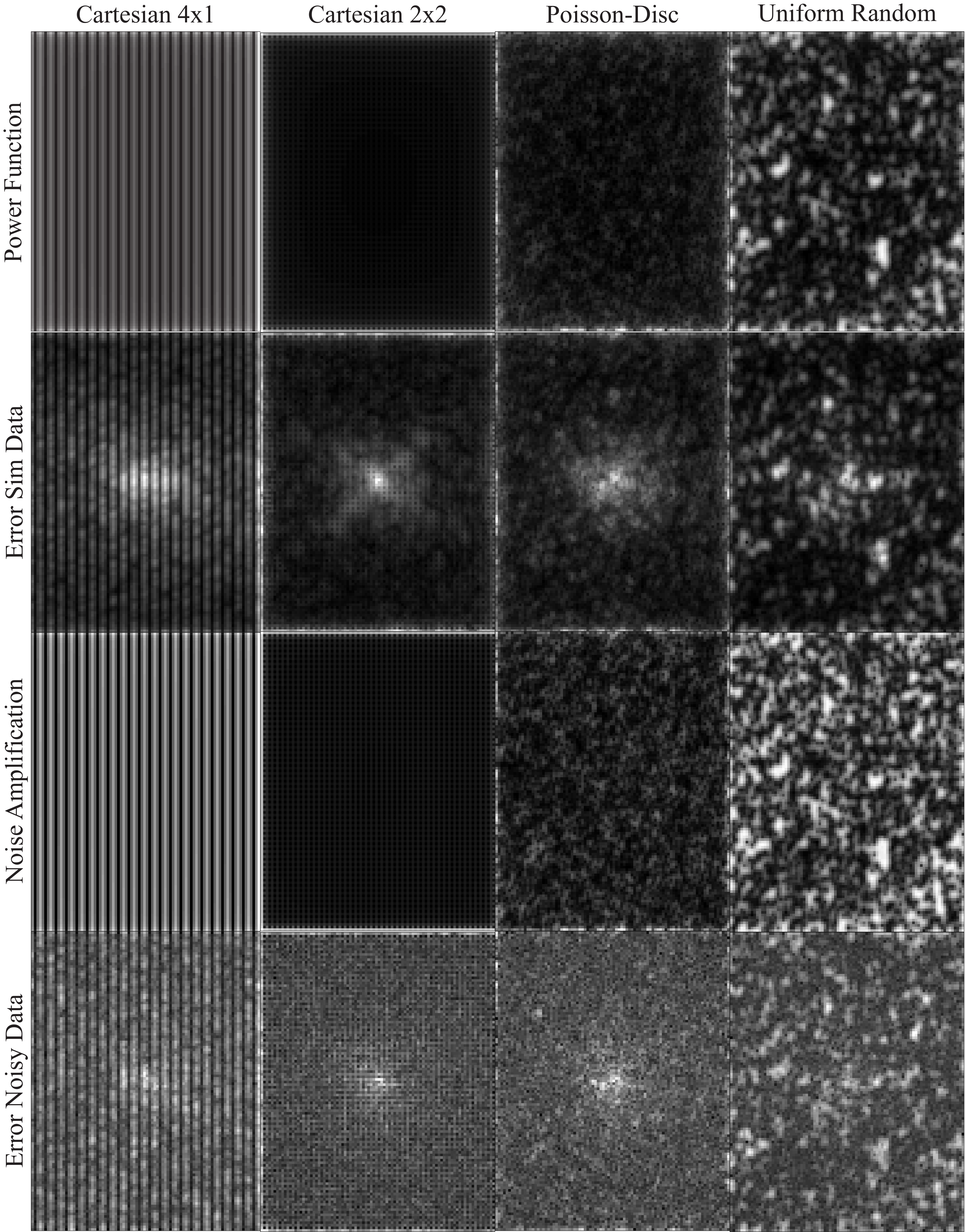}
\caption{Theoretical power function and noise amplification maps as 
well as actual reconstruction errors in k-space for simulated and noisy data.
Because energy of the error
is much higher near the center of k-space, the maps have been raised to a power 
of $1/3$ for improved visualization of their structure.
For this reason, please note that the relative
intensity of the different maps is misleading.
}
\label{fig:KERROR}
\end{center}
\end{figure}

Figure~\ref{fig:SENS} shows the coil image and
the corresponding sensitivity map for all receive channels as well
as the combined image.
Figure~\ref{fig:POWER} shows the combined power function and the Frobenius
norm of the cardinal functions for different sampling patterns.
While for Cartesian $2 \times 2$ and Poisson-disc sampling the power function
is small everywhere inside the sampled region indicating that interpolation
error is small, the situation is different for Cartesian $4 \times 1$ and
uniform random sampling:
Where larger gaps appear in the sampling pattern, the power function
has high values. The power functions themselves are bounded by the diagonal 
elements of the kernel. This bound is approached in regions where the
cardinal functions go to zero, {\it i.e.} far from acquired samples, and
corresponds to a situation where nothing is known about the k-space
value. 
The bound for the combined power functions is $6.5530$ for the kernel 
used here. Consistent with this upper bound, the maximum values 
observed near the boundary in the computed maps are 
$6.4802$ for Cartesian $4\times1$,
$6.3921$ for Cartesian $2\times2$,
$6.3797$ for Poisson-disc, and $6.4277$ for 
uniform random sampling. Computing the maximum in a smaller inner
region of size $305 \times 230$ far from the boundary, the
maximum values are 
$0.6154$, $0.05252$, $0.1087$, and $4.3303$, respectively.
The last number highlights the fact that high values are attained 
even inside the sampled area for uniform random sampling.
While the error bound for Poisson-disc sampling
is twice as large as for the Cartesian $2 \times 2$ pattern, it
is still very small, {\it i.e.} $60 \times$ smaller than the 
maximum which is obtained in unsampled regions.
The reconstruction results (Fig.~\ref{fig:RECO})
for noise-less data confirm that the interpolation error is lower for 
Cartesian and Poisson-disc than for uniform random sampling.
Cartesian $4 \times 1$ performs worse than Cartesian
$2 \times 2$, confirming the notion that
it is usually better to distribute the acceleration along
different phase-encoding directions.
The structure of the error maps in k-space is predicted well 
by the power function for all sampling patterns (Fig.~\ref{fig:KERROR}).
It has to be noted that the power function yields only a worst-case bound
(scaled by the norm of the data)
which depends on the sampling pattern, but not on the actual signal.
In contrast, the actual error values in k-space depend on the energy
distribution of the signal and are much higher in the k-space center
than in the periphery.

In addition to the interpolation error,
noise is amplified during the reconstruction. Assuming Gaussian white
noise, this effect is described by the Frobenius norm of the
cardinal functions. 
In Nyquist-sampled regions, if all channels contribute equally
one would expect a value of $\sqrt{1/N}$  because 
the data from all channels is averaged. Values can be much 
higher in case of undersampling, but can also be lower
for regions very far from acquired samples. This can be seen
in at the boundary of the computed maps shown in Figure~\ref{fig:POWER}. 
In agreement with the higher values
of the Frobenius norm for 
Cartesian $4 \times 1$ and
uniform random sampling, the respective reconstruction
results for the noisy data show much more noise in the reconstructed 
image. Again, the distribution of noise and errors in k-space has 
the same structure as the Frobenius norm of the local reconstruction 
operators and the power function predict (Fig.~\ref{fig:KERROR}).

\begin{figure}
\begin{center}
\includegraphics[width=1\columnwidth]{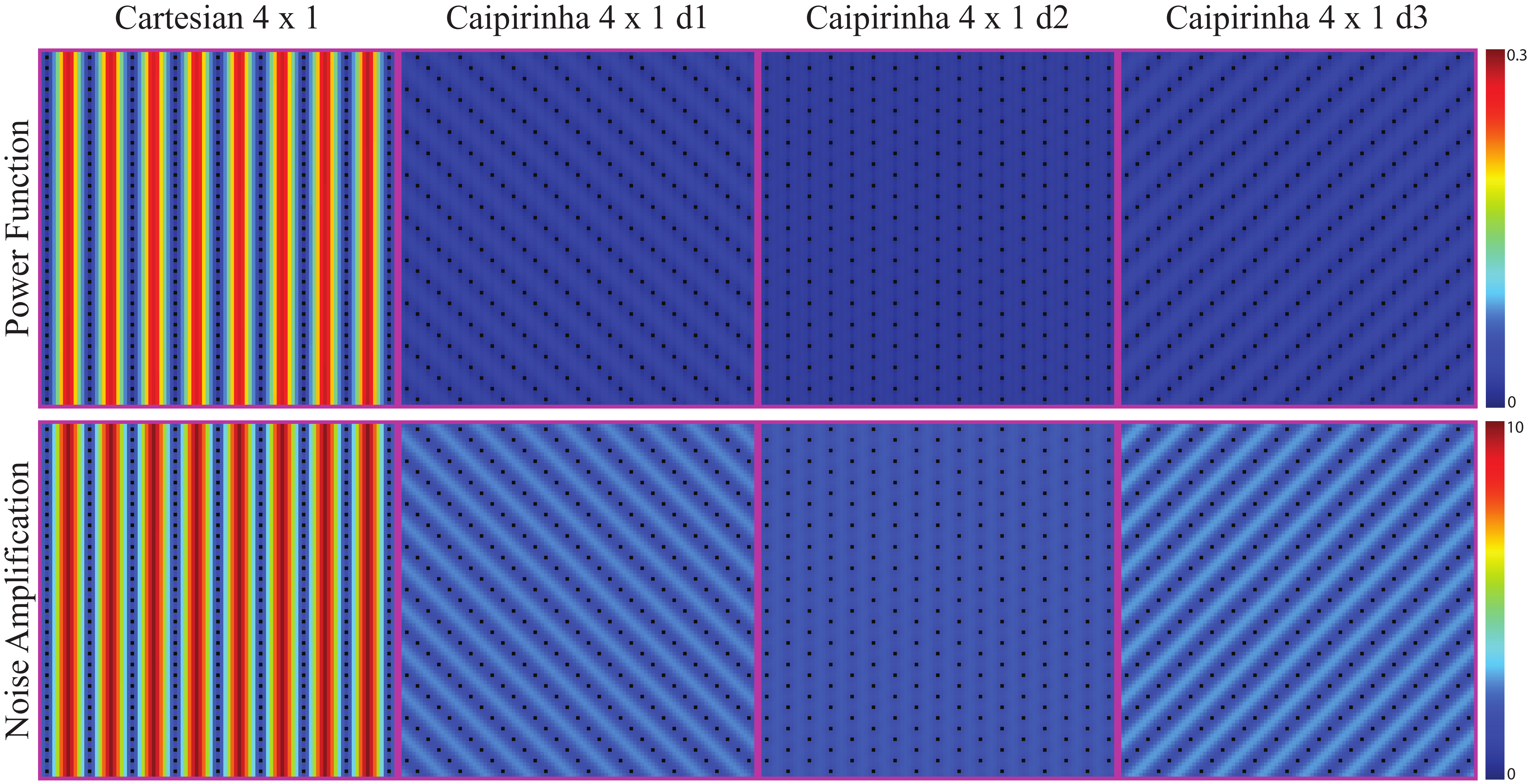}
\caption{
Sampling pattern, combined power function, and local noise amplification 
for Cartesian $4 \times 1$ and three different CAIPIRINHA sampling patterns
with different shifts. Only a part of the full grid is shown.}
\label{fig:CAIPIP}
\end{center}
\end{figure}

\begin{figure}
\begin{center}
\includegraphics[width=1\columnwidth]{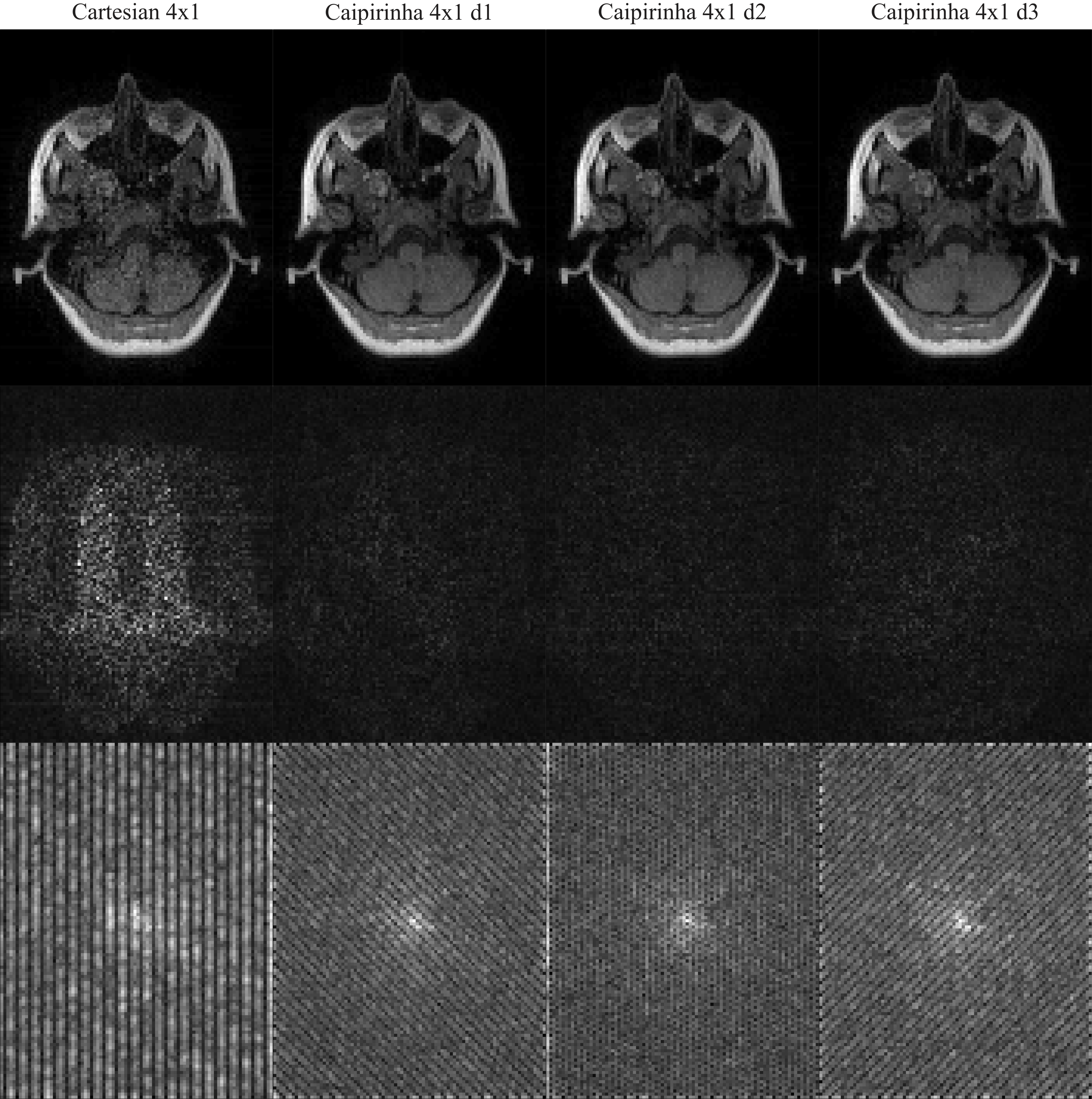}
\caption{
Images reconstructed from noisy data (top) and corresponding error maps in the
image domain (middle) and k-space domain (bottom) for Cartesian $4 \times 1$
sampling and CAIPIRINHA with different shifts. 
The k-space maps have been raised to a power 
of $1/3$ for improved visualization.}
\label{fig:CAIPIR}
\end{center}
\end{figure}

The differences between Cartesian $4 \times 1$ and 
the CAIPIRINHA patterns for the power function and Frobenius
norm of the cardinal functions are shown in Figure~\ref{fig:CAIPIP}.
Essentially distributing the undersampling in both phase-encoding
dimensions, all CAIPIRINHA patterns show much lower values for both
functions than the Cartesian $4 \times 1$ pattern. The CAIPIRINHA
pattern with a shift of two performs slightly better with respect
to noise amplification than the two others. The predictions 
are confirmed in the reconstructions from noisy data and 
corresponding error maps in image and k-space domain (Fig.~\ref{fig:CAIPIR}).

In summary, the results confirm the intuition that 
Cartesian $2 \times 2$ and Poisson-disc sampling yield better
k-space interpolation and less noise
amplification and consequently better image reconstruction than
uniform random sampling. Poisson-disc sampling is
only slightly worse than Cartesian $2 \times 2$ sampling. 
Also as expected, Cartesian $4 \times 1$
performs worse than Cartesian $2 \times 2$ and
all CAIPIRINHA $4 \times 1$ sampling patterns.
The new proposed metrics can predict local
reconstruction errors in k-space for different 
sampling patterns.

\section{Discussion}

\subsection{Parallel Imaging as Approximation in an RKHS}

In this work, it has been shown that
the space of ideal multi-channel k-space signals in MRI is an RKHS. 
As such, it is completely characterized by its kernel, which can 
be derived from the spatial sensitivity profiles of the receive
coils. Based on this result, the connection to approximation 
theory is fully developed.
The interpolation formula (Eq.~\ref{eq:INTERPOL}) allows optimal 
reconstruction from samples at arbitrary locations
in k-space, {\it i.e.} it provides a solution to the reconstruction
problem of Cartesian and non-Cartesian parallel MRI. If samples are 
acquired on arbitrary positions, the kernel (Eq.~\ref{eq:KERNEL})
can be evaluated by non-uniform FFT techniques~\cite{SULLIVAN,JACKSON,BEATTY}.
It has to be acknowledged that solving Equation~\ref{eq:CARDINAL} 
is far too expensive for practical applications. In fact, image
reconstruction using iterative optimization of Equation~\ref{eq:MINI}
still seems to be the best approach due to its efficiency and 
flexibility. This does not mean that Equation~\ref{eq:CARDINAL}
is of purely theoretical interest. While the present study focussed
on numerical exactness, practical algorithms such as GRAPPA~\cite{GRAPPA}
and PARS~\cite{PARS} can be understood as a local approximation of this
formula. Valuable insights can also be expected by analyzing other 
existing methods in this framework. 
For example, the calibration-consistency condition $W x = x$ used 
in SPIRiT is a discrete local version of the reproducing property (Eq.~\ref{eq:REPRO}).
On the other hand, methods which use a nonlinear model to jointly
estimate image content and coil sensitivities~\cite{JSENSE,Uecker08},
or use non-linear regularization terms can not directly be addressed.

\subsection{Error Bounds}

Previous work in parallel imaging uses g-factor maps to quantify 
noise behaviour in the image domain. These maps
can be calculated analytically for periodic sampling patterns 
for SENSE~\cite{SENSE} and GRAPPA~\cite{BREUERGF}. 
For arbitrary sampling patterns g-factor maps can be computed
using Monte-Carlo methods based on full reconstructions~\cite{Robson2008}.
While the g-factor map is a valuable tool to assess noise
in the reconstructed image, it does not offer any direct 
insights into the source of these errors in k-space.

The present work describes new tools
to study approximation error and noise amplification in k-space.
The power functions can be used to predict local approximation errors 
for different sampling patterns, and the noise behaviour can
be analyzed using the Frobenius norm of the cardinal functions. 
This is demonstrated in several experimental examples. 
CAIPIRINHA patterns have been developed to improve
parallel imaging in 3D MRI by shifting samples in each k-space
row by a different amount. Using the power function, it was
directly confirmed that this leads to smaller errors bounds
between samples. In the important combination
of compressed sensing and parallel imaging~\cite{TVMRI,LIU,Lustig2010},
sampling schemes must provide incoherence while optimally exploiting
the information from multiple coils. In this context, Poisson-disc
sampling has been proposed as a replacement for random sampling
based on the idea that the close area 
around a sample should not be sampled again because it is
highly correlated for multiple coils and can be 
recovered with parallel MRI~\cite{PRACTICAL}. 
This intuitive idea could be confirmed for a specific
coil array by comparing the power function
for different sampling schemes. It is noteworthy that
the lowest values for the power function could be achieved
with Cartesian sampling. Although the power function
yields useful error bounds in k-space, it is important
to keep in mind that the optimal choice 
of the sampling scheme may depend on other factors 
such as the structure of the aliasing in the image domain.
A limitation of the present work is that only
a linear reconstruction is considered, while
compressed sensing is a non-linear method.
In compressed sensing, random or Poisson-disc sampling is 
used to produce 
incoherent aliasing which can then be removed using 
sparsity constraints to achieve higher acceleration.

\subsection{Extensions}

In the present study, it was assumed that all channel are always
sampled simultaneously. While this is a reasonable assumption for
data acquisition in MRI, the mathematical theory does not impose
this limitation. In fact, relaxing this condition allows some
interesting extensions. For example, by augmenting the RKHS with a 
uniformly sensitive ``coil'' that collects no data~\cite{BEATTYDVC},
it is possible
to bound point-wise approximation errors of the Fourier transform of 
the single underlying image, ultimately the quantity of interest.
Another possible application is the evaluation of the individual contribution
of samples from different coils, which might be useful in the context of
coil selection and array compression schemes~\cite{ARCOMP,ARCOMP2,Tao2012}.
An interesting application of the new metrics derived in this
work could be the automatic
design of optimal sampling patterns. For example, such techniques
have previously been developed based on a greedy approach using
an analytical formula for the global noise error~\cite{Xu2005},
using simulated annealing using an approximate 
reconstruction~\cite{Gong2013}, or Bayesian experimental
design~\cite{Seeger2010}. In these applications, 
localized error metrics in k-space could be used 
to guide the automatic selection of new sample points.

In this work, regularization has been used only at a small level to
stabilize the numerical computations with the ill-conditioned
kernel matrix. A higher regularization parameter can be used to 
optimize the trade-off between noise amplification and
approximation error. This will be studied in a future work.

Coil sensitivities have to be estimated from experimental
data, for example using ESPIRiT. One important result of ESPIRiT is 
that in some cases of corruption
the coil sensitivities can not be determined uniquely.
In this case, multiple sets of maps~$c^l_j$ for $l=1\dots L$ appear, 
which have to be used simultaneously in the reconstruction.
In the framework described here, this corresponds to
kernels of the form
\begin{align}
	K_{ij}(\xb,\yb) & :=  
	 \int_{\Omega}d\rb \, e^{-2\pi\sqrt{-1}(\xb-\yb) \cdot \rb}  \sum_{l=1}^L c^l_{i}(\rb)\overline{c^l_{j}(\rb)}~.
\end{align}

Finally, it should be noted that the framework of approximation
theory is very general and can be applied to completely
different encoding schemes, {\it e.g.} non-linear gradient
fields or multi-slice excitation~\cite{PATLOC,MULTISLICE}.

\section{Conclusion}

In the present work, parallel MRI has been formulated
as an approximation of vector-valued functions in a RKHS.
This space can be completely characterized by a kernel derived from
the sensitivities of the receive coils. The new formulation provides 
a sound mathematical framework for understanding the reconstruction
process and sample selection, which has been demonstrated by
experimental examples comparing Cartesian, Poisson-disc,
and random sampling.

\section{Acknowledgements}

The authors thank Robert Schaback for helpful discussions. 
This study was supported by
NIH grants P41RR009784 and R01EB009690, 
American Heart Association 12BGIA9660006, Sloan Research Fellowship,
GE Healthcare,
and a National Science Foundation Graduate Research Fellowship.

\section{Appendix}

\subsection{Notation}\label{app:NOTATION}

Important symbols are listed in Table~\ref{tab:NOTATION}.
Bold quantities denote vectors or vector-valued functions.
An upper subscript denotes a relationship to another
quantity, {\it e.g.} $\Rb^{\xb,i}$ is the representer of evaluation
at k-space position $\xb$ for channel $i$. Lower subscripts
always denote discrete indices which select a component of 
a vector-valued function or an element out of a set.
For example, the representer $\Rb^{\xb,i}$ itself is a
vector-valued function, which can be evaluated at a 
specific sample position $\yb_l \in S$ and subscripted to 
obtain the component for a specific channel $j$, which
could then be written as $K^{\xb,i}_j(\yb_l)$.

\subsection{Interpolation}\label{app:EXACT}

The interpolation formula given in Equation~\ref{eq:INTERPOL} computes 
the projection~$\fb^{\parallel}$ 
of any function $\fb \in H$ onto $H_S$ from its samples $\fb(\xb_k)$ 
with $\xb_k \in S$. From the reproducing property~(Eq.~\ref{eq:REPRO}) and
the definition of $H_S$ follows that $\fb^{\perp}({S}) = \{ 0 \}$.
$\fb^{\parallel} \in H_S$ can be interpolated exactly.
This can be shown by expressing an arbitrary function $\fb^{\parallel} \in H_S$ by
its samples $\fb^{\parallel}(\xb_k)$ at $\xb_k \in S$:
\begin{align}
	\fb^{\parallel}(\xb) & = \sum_{l=1}^{|S|} \sum_{j=1}^{N} a_{\xb_l,j} \Rb^{\xb_l,j}(\xb) \\
		& = \sum_{l=1}^{|S|} \sum_{j=1}^{N} a_{\xb_l,j} \sum_{k}^{|S|} \sum_{i=1}^N K_{ij}(\xb_k, \xb_l) \ub^{k,i}(\xb) \\
		& =\sum_{k=1}^{|S|} \sum_{i=1}^N \sum_{l=1}^{|S|} \sum_{j=1}^{N} a_{\xb_l,j}  K_{ij}(\xb_k, \xb_l) \ub^{k,i}(\xb) \\
	&  = \sum_{k=1}^{|S|} \sum_{i=1}^N f^{\parallel}_i(\xb_k) \ub^{k,i}(\xb) 
\end{align}

\subsection{Error Bounds}\label{app:BOUND}

For functions in $H$, an point-wise error bound can be computed:
\begin{align}
	e_n(\xb) & = |\fb_n(\xb) - \hfb_n(\xb) | \\
	& = \bigg| \fb_n(\xb) - \sum_{k=1}^{|S|} \sum_{i=1}^N f_i(\xb_k) u^{k,i}_n(\xb) \bigg| \\
	   & = \bigg| \Big\langle \Rb^{\xb,n} - \sum_{k=1}^{|S|} \sum_{i=1}^N \Rb^{\xb_k,i} u^{k,i}_n(\xb), \fb \Big\rangle_H \bigg| \\
	 & \leq \| \fb \|_H \underbrace{\bigg\| \Rb^{\xb,n} - \sum_{k=1}^{|S|} \sum_{i=1}^N \Rb^{\xb_k,i} u^{k,i}_n(\xb) \bigg\|_H}_{P_n(\xb)}
\end{align}
Where $P_n$ is the $n$-th component of the power function. Using
the reproducing property on the kernel itself
$\left< \Rb^{\xb,i}, \Rb^{\yb, j} \right>_H = K_{ij}(\xb, \yb)$, it
can be expressed as:
\begin{align}
 	P_n^2(\xb) 
=  K_{nn}(\xb, \xb) - 2 {\mathcal Re} \sum_{k=1}^{|S|} \sum_{i=1}^N K_{in}(\xb_k, \xb) u_n^{ik}(\xb) \\
	\qquad + \sum_{l=1}^{|S|} \sum_{k=1}^{|S|} \sum_{i=1}^N \sum_{j=1}^N K_{ij}(\xb_k, \xb_l) u_n^{ik}(\xb) \overline{u_n^{jl}(\xb)}
\end{align}
For an unregularized least-squares reconstruction the interpolation error is orthogonal to the interpolant, {\it i.e.}
$<\fb - \hat \fb, \hat \fb >_H = 0$. In this case, the power function can be simplified to:
\begin{align}
 	P_n^2(\xb) =   K_{nn}(\xb, \xb) - \sum_{l=1}^{|S|} \sum_{j=1}^N K_{nj}(\xb, \xb_l) \overline{u_n^{jl}(\xb)}
\end{align}

\newpage
\bibliography{rkhs}

\newpage

\end{document}